\documentclass[]{elsart}
\usepackage{graphics}
\usepackage{times}
\usepackage{latexsym}

\begin{document}
\begin{frontmatter}

\title{
Are there universal freeze-out parameters for central 158$A$GeV Pb+Pb 
collisions?
}

\author{T.Peitzmann\thanksref{email}} 
\address{University of M{\"u}nster, D-48149 M{\"u}nster, Germany}
\thanks[email]{email: peitzmann@ikp.uni-muenster.de}
\begin{abstract}
	Earlier attempts to extract parameters of kinetic freeze-out in 
	Pb+Pb collisions at 158$A$GeV are critically reviewed. Many of these 
	analyses have used approximations which have significant impact on the 
	extracted parameters. Simple estimates are 
	obtained which attempt to avoid the most critical approximations. It 
	is pointed out that constraints based on pion interferometry are 
	less reliable than those from momentum spectra. 
	A universal set of freeze-out parameters from 
	transverse mass spectra would require 
	$T \ge 135 \, \mathrm{MeV}$ 
	and $\left\langle \beta_{T} \right\rangle \le 0.35$.
\end{abstract}

\end{frontmatter}
\section{Introduction}

Experiments with ultrarelativistic heavy ions at the CERN SPS which 
are on the quest for a new state of strongly interacting matter have 
obtained a wealth of information during the last years (see e.g. 
\cite{qm97,qm99}). While there are several interesting non-trivial 
observations in these experiments, our understanding of the evolution 
of the reactions is far from complete. However, for an interpretation of 
the promising possible signatures of the quark-gluon plasma a better 
understanding of the reaction dynamics is mandatory.

Hydrodynamical models or simple parameterizations have been used 
as tools to help understanding some of the issues in heavy ion 
reactions. While even the more complex versions of such models are 
believed to be crude approximations, it is worthwhile to use 
them and also the simpler parameterizations to test strongly different 
scenarios and provide first guidelines on the possible dynamics, if 
the reaction mechanisms are not too far away from the concept of local 
thermal equilibrium. One of the key questions in this context is the 
partition of the excitation energy into collective and thermal (i.e. random) 
motion of the produced particles. This is mostly characterized by the 
average values of the temperature and the radial flow velocity of the 
finally observed hadrons, the so-called kinetic freeze-out parameters.

Still, in such comparisons one has to be cautious to 
\begin{enumerate}
	\item  use only those experimental results which are most likely too 
	share the same equilibrium conditions and

	\item  avoid approximations that are wrong even in the most 
	optimistic cases.
\end{enumerate}
In the following I will study some of the experimental results on 
momentum spectra and interferometry
that have been used in applications of these hydrodynamical models 
and check their implications for the estimates of freeze-out 
parameters. I will first use estimates based on 
effective slopes of transverse momentum spectra for different 
particle species and compare them with the results of hydrodynamical 
parameterizations of neutral pion spectra. I will then turn to the 
interpretation of pion interferometry results. There the momentum 
dependence of effective radii extracted from fits to two-particle 
correlation functions is used to extract information about temperature 
and flow. As the method to obtain this information is much more 
indirect, it is naturally more complicated to interpret the 
corresponding results. I will try to demonstrate that some of the 
recipes established should be critically revisited.

\section{Analysis}

\subsection{Inverse Slopes}
The transverse momentum spectra of hadrons are believed to be 
frozen in a relatively late stage of the nuclear collision. If there 
is local equilibration, the elementary fireball volumes 
can be associated with local temperatures. In a finite 
system there will necessarily be hydrodynamic motion because of 
the pressure gradient. Therefore, at freeze-out the system may be 
characterized by local temperatures and transverse flow velocities 
which will determine the shape of the transverse momentum 
spectra.\footnote{There is an additional dependence of the shape on 
particle production by resonance decay which I will not discuss in 
detail here.} If the composition of the system is relatively 
homogeneous, the freeze-out may occur at a relatively similar 
temperature throughout the whole volume. It is therefore reasonable 
to speak of one general freeze-out temperature. One should keep in 
mind, however, that this is a crude approximation: It is more a 
question of how different all the local temperatures are and for which 
particle species they are relevant. This simplification may be 
used for the discussion of momentum spectra, where a superposition of 
distributions with different temperatures may well be approximated by 
one distributions with an average temperature. This simplification is 
much more crucial in the case of pion interferometry, as I will 
discuss below.

The general idea in using different particle species to estimate 
these parameters relies on the fact that in a system with a 
temperature and a collective flow velocity, heavier particles will 
acquire higher momenta from the shared velocity because of their 
larger mass and will exhibit flatter momentum spectra, i.e. larger 
inverse slopes.
In \cite{NA44b} the authors have presented inverse slopes $T_{\rm eff}$ of 
transverse mass spectra of different particle species and used those 
to estimate a freeze-out temperature and flow velocity from the 
relation:
\begin{equation}
\label{eq:na44}
    T_{\rm eff} = T + m \cdot 
    \left\langle \beta_{T} \right\rangle^2.
\end{equation}
They obtain a freeze-out temperature of $T \approx 140 \, \mathrm{MeV}$. 
This analysis has been criticised as neglecting relativity 
\cite{na44:nix:98,nix:98}, the authors of the latter publications present a better 
estimate of the slopes:
\begin{equation}
\label{eq:nix}
    T_{\rm eff} = \sqrt{1 - \left\langle \beta_{T} \right\rangle^2}
	\frac{T}{1 - \left\langle \beta_{T} \right\rangle
	\sqrt{1 + m^2/p_{T}^2}}.
\end{equation}
Then the limit of zero particle mass:
\begin{equation}
\label{eq:nix2}
    T_{\rm eff} = T \sqrt{\frac{1 + \left\langle \beta_{T} \right\rangle}
	{1 - \left\langle \beta_{T} \right\rangle}}
\end{equation}
is used to reestimate the freeze-out temperature. They obtain $T \approx 
90 \, \mathrm{MeV}$ using the data of \cite{NA44b}. 
Both groups use also somewhat more sophisticated models to obtain 
freeze-out parameters which support their simple temperature estimates. 

I would like to demonstrate the ambiguities in such estimates 
by using the correct relativistic 
formula \ref{eq:nix} to calculate freeze-out temperatures from the 
same slope parameters. For this purpose I have to assume 
$p_{T} = \left\langle p_{T} \right\rangle$. This is of course an approximation 
which should however not be as severe as those used in \cite{NA44b} and 
\cite{na44:nix:98,nix:98}. The slopes and the values 
of $\left\langle p_{T} \right\rangle$ used are given in table \ref{tbl1}. 
It turns out that not all three particle species investigated can be 
described with the same set of parameters. Comparing positive kaons 
and protons yields $T = 139 \, \mathrm{MeV}$ and 
$\left\langle \beta_{T} \right\rangle = 0.33$, while for negative 
kaons and antiprotons one gets $T = 152 \, \mathrm{MeV}$ and 
$\left\langle \beta_{T} \right\rangle = 0.28$. However, estimates 
using pions and (anti)protons yield $T \approx 100 \, \mathrm{MeV}$ and 
$\left\langle \beta_{T} \right\rangle \approx 0.41$ and those for 
pions and kaons yield $T \approx 84 \, \mathrm{MeV}$ and 
$\left\langle \beta_{T} \right\rangle \approx 0.51$.

Although the authors of \cite{NA44b} exclude the region of $m_{T} - 
m < 0.25 \, \mathrm{GeV}/c$ from the fit for pions, it is still very 
likely that the slopes for pions are influenced by resonance decays 
which would result in a lowering of the inverse slope and 
thus implicitly lower the extracted freeze-out temperature. Kaons and 
protons should be relatively free from such influence, so that the 
estimate using these particle species should be more reliable. The 
temperature $T = 140 - 150 \, \mathrm{MeV}$ is also in line with the 
hydrodynamical fit analysis in \cite{NA44b}. 

\begin{figure}[tb]
	\centering
	\includegraphics{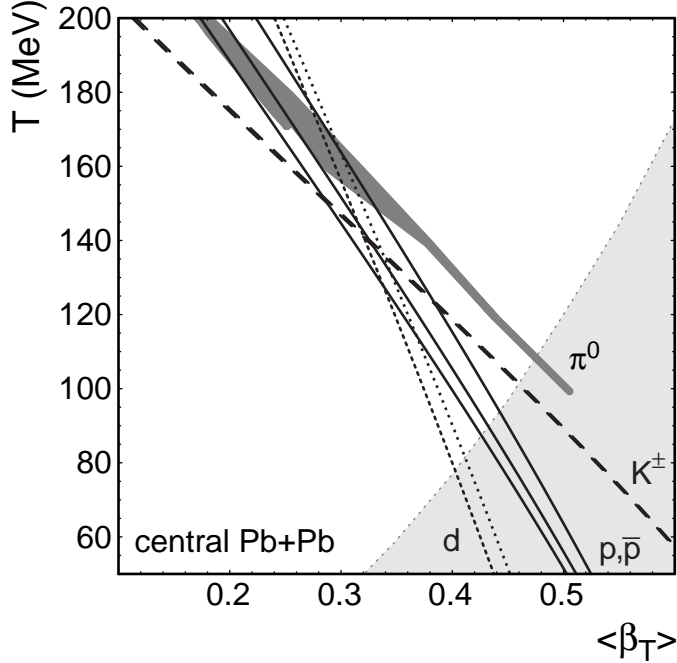}
	\caption{Relations between temperature and flow velocity extracted 
	from hydrodynamical parameterizations of neutral pion spectra from 
	WA98 (grey band) and from slope parameters of proton and antiproton 
	(NA44 and NA49, solid lines), kaon (NA44, dashed lines) and deuteron spectra 
	(NA44 and NA49, dotted lines). The light grey area shows the allowed 
	region from pion interferometry analysis by NA49.
	(For details see text.)}
	\label{fig1}
\end{figure}

The detailed analysis in 
\cite{na44:nix:98} which obtains lower freeze-out temperatures uses 
additional information, namely the normalization of the spectra and 
two-particle correlations. It is known from other analyses (see e.g. 
\cite{na49hydro}) that two-particle correlation which can provide 
constraints on $\left\langle \beta_{T} \right\rangle^2/T$ within 
hydrodynamical models effectively 
gives very little information on the temperature but constrains mostly 
the flow velocity (see light grey area in figure~\ref{fig1}). 
On the other hand the normalization of spectra of 
different particle species is related to \emph{chemical freeze-out}, 
while the spectral shape is influenced by \emph{kinetic 
freeze-out}.\footnote{There is an influence of the chemical 
temperature on the spectral shape via the resonance decays -- this 
should however be relevant only for pions.} 
To assume that these two corresponding temperatures are the same, as 
was apparently done in \cite{na44:nix:98}, is very likely 
wrong.\footnote{A similar assumption is used in the analysis in 
\cite{ster:budalund1,ster:budalund2} which will not be discussed here.}

From the large pion-nucleon cross section one would assume that these 
two particle species are most likely to share the same freeze-out 
conditions. However, the influence of resonance decays in the pion 
spectrum forbids the simple use of slope parameters to estimate 
freeze-out conditions. To perform such a comparison the slope 
parameters for heavier particles will be used together with a full 
hydrodynamical parameterization of neutral pion spectra as described 
in \cite{wa98:hydro:99}. The hydrodynamical computer program 
described in \cite{wiedemann96} has been modified to 
describe neutral pions and accommodate a Woods-Saxon density profile 
in addition to the default Gaussian profile
and was used to obtain best parameter sets for $T$ and 
$\left\langle \beta_{T} \right\rangle$. The $2\sigma$-allowed values 
obtained for different profiles (Gaussian and Woods-Saxon of different 
diffuseness) are shown 
in figure \ref{fig1} as a grey band. Among these fits, those obtained 
with large diffuseness (as e.g. the Gaussian) tend to yield smaller 
flow velocities for the same temperature but also constrain the 
allowed parameters to the high $T$ region \cite{wa98:hydro:99}.
In addition, the slope 
parameters for kaons, protons, antiprotons and deuterons from NA44 and 
NA49 as summarized in table \ref{tbl1} 
have been used with the help of equation \ref{eq:nix} to obtain 
similar relations between $T$ and 
$\left\langle \beta_{T} \right\rangle$. 
All these estimates are also displayed in 
figure \ref{fig1}.

The NA49 collaboration has also published freeze-out constraints 
obtained from negative hadron and deuteron transverse mass spectra 
\cite{na49hydro}. The negative hadrons are dominated by pions which 
should allow to use them in hydrodynamical descriptions assuming the 
pion mass, however, the analysis was performed using a 
non-relativistic approximation and ignoring resonance production. The deuteron 
spectra are not affected by resonances but still suffer from the non-relativistic 
approximation. The 
corresponding parameters are therefore not reliable. In fact, the 
estimate from the slope included in figure \ref{fig1} differs 
considerably from the constraints in \cite{na49hydro}, esp. at large 
velocities. While the curves for kaons, protons and deuterons seem to 
converge around $T = 140 \, \mathrm{MeV}$ and 
$\left\langle \beta_{T} \right\rangle = 0.33$, the neutral pion 
results have a further bias to somewhat higher temperatures. One 
should note that the analysis of the pion spectra differs from the 
others, as 
\begin{enumerate}
	\item  a full parameterization including the effects of resonance 
	decays has been performed compared to using an effective slope 
	together with an approximate relation connecting $T$ and 
	$\left\langle \beta_{T} \right\rangle$ and

	\item  they cover a different $m_{T}$ range -- the high $m_{T}$ tail 
	of the neutral pion spectrum has also been used as an upper limit as 
	described in \cite{wa98:hydro:99}.
\end{enumerate}
Still this figure suggests that a common parameter set for all 
the momentum spectra considered here can only be obtained for 
$T \ge 135 \, \mathrm{MeV}$ and $\left\langle \beta_{T} \right\rangle \le 
0.35$. This appears to contradict constraints obtained from pion 
interferometry measurements \cite{na49hydro} (see light grey area in figure 
\ref{fig1}), which favor larger 
transverse flow velocities.

\subsection{Interferometry}
It is well known that correlations 
between position and momentum space modify the radii measured in pion 
interferometry. One of the most interesting examples of such a 
correlation is collective flow \cite{mak88,sin89}, 
both longitudinal and transverse, 
although there are other examples, like a source with local 
temperatures $T = T(x^{\mu})$, as I will discuss below. In the 
idealized case of pure collective motion (no thermal contribution) 
only the source element moving towards the detector will be measured, 
so the radius extracted from interferometry will correspond to this 
small elementary volume. If there is additional thermal motion, the 
probability of a particle to still be measured while being radiated 
from another source element 
at another space point and directed slightly away from the detector 
will increase. The effective source size will therefore decrease, if 
the amount of collective motion relative to thermal motion is large, 
and vice versa. In addition, this effect is a function of the 
momentum of the particle, and in most scenarios the effective radii 
will decrease with increasing transverse mass of the particle.

Most quantitative studies of these effects start from the Cooper-Frye 
formula \cite{cooperfrye} for the momentum distribution of the source:
\begin{equation}
	E\frac{d^3N}{dp^3} = \frac{2J+1}{(2\pi)^3}
	\int_{\Sigma}d^3\sigma_{\mu} \frac{p^{\mu}}{\exp\left\{ \left[ p_{\mu} 
	v^{\mu} - \mu \right]/ T \right\} \mp 1}.
    \protect\label{eq:cf1}
\end{equation}
In this formula the factor $p^{\mu}$ in the numerator and the exponent 
$p_{\mu}v^{\mu}-\mu$ originate from the Lorentz transformation with 
$v^{\mu} = v^{\mu}(x^{\mu})$ being the local four-velocity of the 
source elements. In general also $T$ is again a local variable. $\mu$ 
is the chemical potential and the $\mp$ signs are relevant for bosons 
and fermions, resp.

There has been substantial work on extracting expressions for 
observables in interferometry, the most systematic approach using a 
parameterization of the source distribution and a saddle point 
approximation to extract approximate analytic relations for the 
effective radii \cite{Chap95}.
A relatively general result is obtained using a source distribution 
expressed as:
\begin{equation}
   S(x,K) = \frac{m_t\, {\rm ch}(\eta-Y)}
            {(2\pi)^3 \sqrt{2\pi(\Delta \tau)^2}}
      \exp \left[- \frac{K{\cdot}u}{T}
                 - \frac{x^2+y^2}{2R_{G}^2}
                 - \frac{(\tau-\tau_0)^2}{2(\Delta \tau)^2}
                 - \frac{\eta^2}{2(\Delta \eta)^2}
           \right] .
\label{eq:heinz11a}
\end{equation}
Gaussians are used for the space-time and rapidity distributions, 
the momentum spectrum is approximated as a Boltzmann 
distribution:
\begin{equation}
   E\frac{d^3N}{dp^3} \propto \exp \left(- \frac{K{\cdot}u}{T}
            \right) .
\label{eq:Boltz1}
\end{equation}
For a model with transverse expansion, 
the authors obtain approximate relations for the 
transverse and longitudinal radii:
\begin{eqnarray}
R_{T} & \approx & R_{G}\left[1+\frac{m_T\beta_{T}^2}{T}{\rm ch}
(\bar{\eta}-Y)\right]^{-1/2} \label{eq:heinzrperp} \\
R_{L} & \approx &    \tau_0\left[\frac{m_T}{T}{\rm ch}(\bar{\eta}-Y)
   -\frac{1}{{\rm ch}^2(\bar{\eta}-Y)} +\frac{1}{(\Delta\eta)^2}
   \right]^{-1/2} . \label{eq:heinzrlong}
\end{eqnarray}
For a boost invariant longitudinally expanding source the expressions 
are more simple, e.g. for the longitudinal radius one obtains: 
\begin{equation}
  R_{L} \approx \tau_0 \sqrt{\frac{T}{m_{T}}} .
\label{eq:heinzrboost}
\end{equation}
As it turns out \cite{na49hydro} that the source rapidity and the pair 
rapidity are similar $\bar{\eta} \approx Y$, also the expression for 
the transverse radius may be simplified:
\begin{equation}
R_{T} \approx R_{G} \left[1 + \frac{\beta_{T}^2}{T} \cdot m_{T} 
 \right] ^{-1/2}.
\label{eq:heinzrperp2}
\end{equation}
These two latter equations are mostly used to extract information 
from experimental two-particle correlation functions. There are a 
number of assumptions which have to be made to obtain equations 
\ref{eq:heinzrperp} and \ref{eq:heinzrlong} and even more for 
equations \ref{eq:heinzrboost} and \ref{eq:heinzrperp2}. Some of 
these are discussed in \cite{Chap95}, however there are further 
assumptions which I will try to concentrate on here.

To be able to use equation \ref{eq:heinzrperp2} to extract 
$\beta_{T}^2/T$ as in \cite{na49hydro}, one of the 
major assumptions is that the parameter $R_{G}$, which contains 
information on the geometrical size of the source, does not 
depend on the transverse momentum of the emitted particles. This is 
not necessarily correct. In general, there will not be 
a universal value for the temperature, it will be a local parameter 
depending on the position of the source element 
with respect to the center or the surface as discussed above.

This induces an additional correlation between momentum and 
coordinate space. If the variations in $T$ are very 
strong, this would of course invalidate all the attempts to extract 
\emph{the} freeze-out temperature. If the variations are 
relatively weak, it may still be reasonable to use one effective 
temperature. It may e.g. be used to characterize momentum spectra, as 
for those the spatial distribution and 
therefore also these correlations are not directly relevant.

\begin{figure}[bt]
	\centering
	\includegraphics{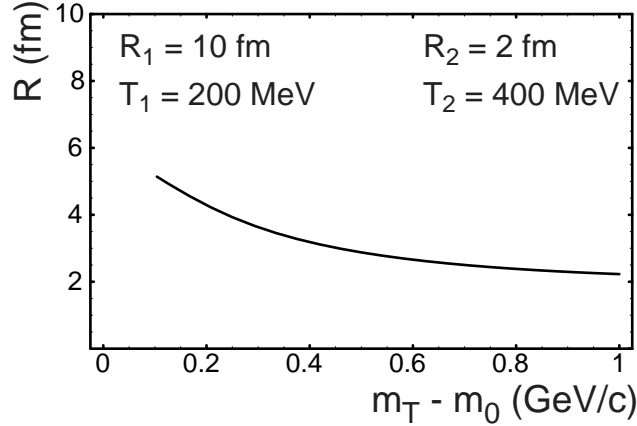}
	\caption{Effective source size as a function of the transverse mass 
	of the particle for a superposition of two different sources (see 
	text).}
	\label{fig2}
\end{figure}

However, they are crucial to the analysis of pion interferometry, 
especially to the above mentioned method of extracting $\beta_{T}^2/T$. 
Consider e.g. that a certain fraction of pions is emitted before 
freeze-out, from the surface of the source, and that this fraction 
increases with increasing momentum. One might try to describe this with a 
superposition of two sources. Just as one example one may consider 
the following -- arbitrary -- scenario: One source of effective size 
(Gaussian) $R_{1} = 10 \, \mathrm{fm}$ and temperature $T_{1} = 200 
MeV$ and a second with $R_{2} = 2 \, \mathrm{fm}$ and $T_{2} = 400 
MeV$.\footnote{For simplicity exponential distributions in transverse 
mass are assumed.} If one investigates produced particles in different intervals 
of transverse mass, the relative contributions of these two sources 
will change, and implicitly the effective source size as seen in 
interferometry. This is shown in figure \ref{fig2}, where the 
effective source size -- again approximated by a Gaussian -- is shown 
as a function of the transverse mass for the case where the two sources 
discussed above both contribute the same fraction to the total 
multiplicity of particles. It can be seen easily that the effective 
source size decreases significantly with increasing transverse mass. 
Of course, this scenario has not been derived from physical arguments 
here, so it should only serve as an illustration of possible effects of 
such correlations between momentum and coordinate space. How large 
such effects might be would require a detailed model calculation 
which is not the scope of this paper. One should only note that 
in general $R_{G}$ in equation \ref{eq:heinzrperp} is expected to be a 
function of $m_{T}$ within the model used. Values of $\beta_{T}^2/T$ 
should therefore be considered as upper limits unless more detailed 
model comparisons are performed. Pion 
interferometry is much more sensitive to such effects than momentum 
spectra.

\begin{figure}[tb]
	\centering
	\includegraphics{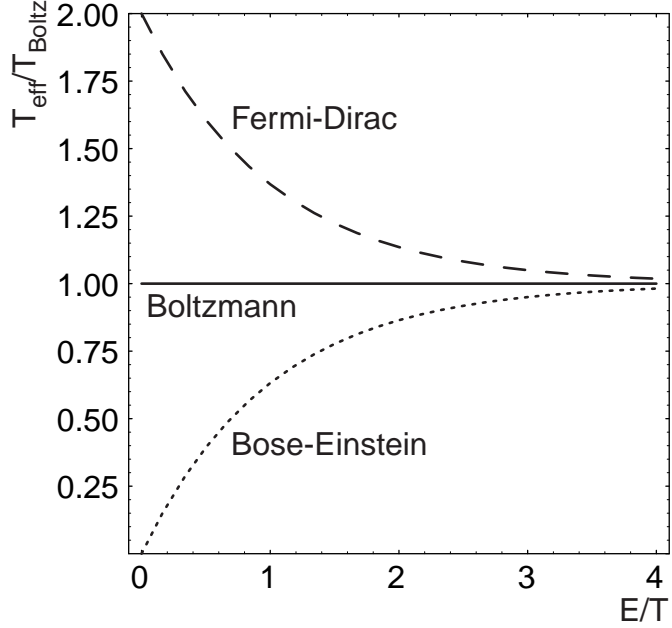}
	\caption{Inverse slopes (equation \ref{eq:invslope}) 
	relative to pure exponential slopes 
	as a function of energy for different 
	distribution functions.}
	\label{fig3}
\end{figure}

A more fundamental caveat to the use of the $m_{T}$ dependence of HBT to 
estimate freeze-out parameters relates to another approximation used in 
extracting the analytical formula which are applied in most of the 
analyses. Specifically, the authors of \cite{Chap95} use the Boltzmann 
approximation to 
the distribution function. While this may be a good approximation for 
most of the pions used in experimental HBT analyses, it is well known 
that the deviations to the true Bose-Einstein distributions are 
crucial at low momenta. This is illustrated in figure \ref{fig3}, 
where the inverse slopes
\begin{equation}
	T_{eff} = - \left[ \frac{d}{d E} \log f(E) \right]^{-1}
	\label{eq:invslope}
\end{equation}
of Bose-Einstein and Fermi-Dirac distributions 
\begin{equation}
   E\frac{d^3N}{dp^3} \propto \frac{1}{\exp \left(- \frac{E}{T}
            \right) \mp 1} .
	\label{eq:befd}
\end{equation}
are shown. Strong deviations from Boltzmann behaviour are seen for 
both distributions for $E/T < 1$. 

\begin{figure}[bt]
	\centering
	\includegraphics{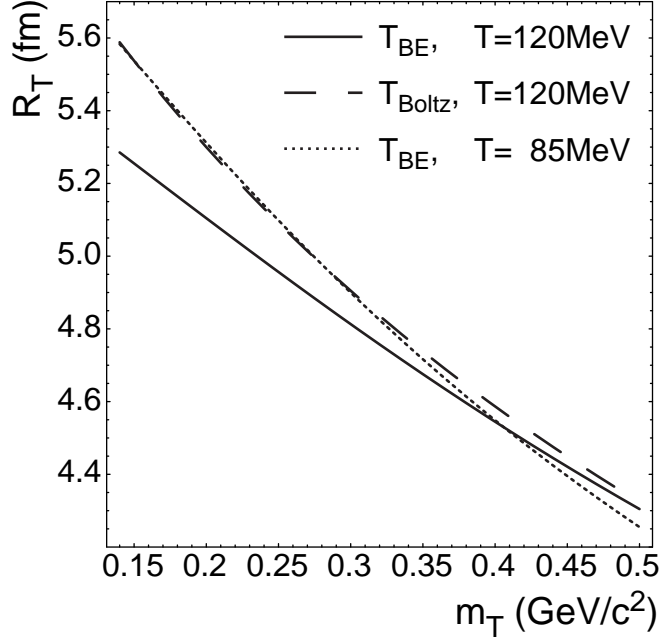}
	\caption{Transverse radii for pion interferometry
	as a function of transverse mass 
	according to equation \protect\ref{eq:heinzrperp2}. The dashed line 
	uses the Boltzmann temperature with values of $T = 120 \, \mathrm{MeV}$ 
	and $\left\langle \beta_{T} \right\rangle \le 0.55$, following the 
	results of \protect\cite{na49hydro}. The solid line uses the same 
	parameters with a correction for the effective slope of a 
	Bose-Einstein distribution. The same correction is used for the 
	dotted line but $T = 85 \, \mathrm{MeV}$ is assumed.}
	\label{fig4}
\end{figure}

As in the analyses in question most of the $m_{T}$ dependence is seen 
at relatively low momenta, one should carefully 
reinvestigate the interpretation of these data with the correct 
statistical distributions. The trend of such a corrected analysis may 
be guessed from the behaviour of the slopes. Using the crude 
assumption that the BE-distributions may be locally approximated by a 
Boltzmann distribution, one might then use the formalism developed 
in \cite{Chap95} to estimate the $m_{T}$ dependence. However, now the 
effective slope for a Bose-Einstein distribution ($T_{BE}$) as shown 
in figure \ref{fig3} should be used instead 
of the Boltzmann temperature $T$. In figure \ref{fig4} the transverse 
radii as a function of transverse mass are shown for three different 
estimates: 
\begin{enumerate}
	\item  Following the formalism as in \cite{na49hydro}, i.e. using the 
	Boltzmann-like temperature in equation \ref{eq:heinzrperp2} and the 
	results obtained in the article, i.e. $T = 120 \, \mathrm{MeV}$, 
	$\left\langle \beta_{T} \right\rangle = 0.55$ and 
	$R_{G} = 6.5 \, \mathrm{fm}$,

	\item  substituting the temperature by the corresponding effective 
	slope of a Bose-Einstein distribution, but using the same parameters 
	as above and

	\item  substituting the temperature by the corresponding effective 
	slope of a Bose-Einstein distribution, but now using 
	$T = 85 \, \mathrm{MeV}$, 
	$\left\langle \beta_{T} \right\rangle = 0.55$ and 
	$R_{G} = 7.1 \, \mathrm{fm}$.
\end{enumerate}
It can be seen from this example that cases 1 and 2 show significant 
differences, and that for a rough agreement with 1 the temperature in 
the corrected version has to be modified from 120~MeV to 85~MeV as in 
case 3.

In this context it is interesting to investigate also results for 
proton source sizes. While the mechanism generating the correlation 
of particle pairs is different compared to pions -- a strong 
interaction quasi-resonance instead of Bose-Einstein statistics, the 
picture of effective source sizes modified by the interplay of random 
thermal and collective hydrodynamical motion remains valid. People 
have therefore applied similar ideas to pp-correlation radii. The 
experiments NA49 \cite{na49pp} and NA44 \cite{na44pp} have measured 
proton correlations at midrapidity in Pb+Pb collisions - they obtain 
Gaussian radii in the range of $R_{pp} = 3.5-4.0 \, \mathrm{fm}$ 
which are considerably smaller that those extracted from pion 
interferometry ($R_{\pi\pi} = 6-7 \, \mathrm{fm}$). Similar values have 
been estimated by applying coalescence models to measured proton and 
deuteron multiplicities \cite{na44pp}. The smaller effective radii 
have been 
attributed to the much larger transverse mass of the protons, 
if a scaling with $1/\sqrt{m_{T}}$ which 
may be derived as an approximation from equations \ref{eq:heinzrboost} 
and \ref{eq:heinzrperp2} is used for both protons and pions. 

From the above argument on BE-distributions it is obvious that 
equivalent modifications are important here. However, the effective slope for a 
FD-distribution is larger at low momenta. In addition, the exponent 
of the FD-distribution contains a non-vanishing chemical potential 
which makes it much more likely that the exponent is small and that
modifications are important. 

Actually, another important argument is relevant here. 
If, as commonly believed, chemical freeze-out occurs relatively early before 
kinetic freeze-out, the relevant picture for the latter is the 
canonical ensemble, as the particle number for each species is fixed, 
so that there is no chemical potential as a parameter. The relevant 
energy in the statistical distributions is the kinetic energy and not 
the total energy, as no new particles are produced and 
the ground state energy of each particle is the 
rest mass. The Cooper-Frye formula
should then be changed to:
\begin{equation}
	E\frac{d^3N}{dp^3} = \frac{2J+1}{(2\pi)^3}
	\int_{\Sigma}d^3\sigma_{\mu} \frac{p^{\mu}}{\exp\left\{ \left[ p_{\mu} 
	v^{\mu} - m_{0} \right]/ T \right\} \mp 1}.
    \protect\label{eq:cf2}
\end{equation}
The major change for the transverse degrees of freedom would be to use 
$m_{T}-m_{0}$ instead of $m_{T}$. This should hold for both bosons and 
fermions, but while it might have only small effects on the pions, it 
will be dramatic for protons because of their large mass.
It is beyond the scope of this paper to estimate the 
effect of such modifications on the $m_{T}$ spectra in general or, more 
specific, on the $m_{T}$ dependence of proton 
source sizes. However, it is also premature to conclude that proton and 
pion source sizes are compatible without performing detailed 
calculations. As the values of $m_{T}-m_{0}$ for protons and pions 
are roughly similar in the experiment, 
no difference in the radii would be expected 
between protons and pions from their transverse masses. This would 
however contradict the measured source sizes.

\section{Conclusion}

I have tried to illustrate above that the freeze-out parameters in 
ultrarelativistic heavy ion collision are not well established 
quantitatively. One should be careful to use only data sets which very 
likely share the same conditions. Also, the used approximations should 
be thoroughly investigated.
The simple re-analysis of some of the available data performed here 
indicates the following:
\begin{itemize}
	\item  Transverse momentum spectra of deuterons, protons, kaons and pions 
	seem to indicate that compatible 
	parameter sets can be obtained only for $T \ge 135 \, \mathrm{MeV}$ 
	and $\left\langle \beta_{T} \right\rangle \le 0.35$.

	\item  The $m_{T}$ dependence of interferometry radii has to 
	be reinvestigated using true statistical 
	distributions instead of Boltzmann approximations at low momenta. 
	This would also require to study the role of modifications related 
	to the change at chemical 
	freeze-out to a system with fixed particle number. 

	\item  Other possible sources of correlations between momentum 
	and coordinate space may cause further alterations of the 
	$m_{T}$ dependence of interferometry radii.
	
	\item  The difference in pion and proton source sizes is very likely 
	not explained by the difference in transverse mass of the two 
	particles.
\end{itemize}

The three last observations shed doubt on the use of pion 
interferometry as the main ingredient to obtain constraints on the 
amount of transverse flow at freeze-out. A more thorough analysis of 
the pion interferometry results is necessary to clarify the situation.

Two possible scenarios may be envisaged which explain the data 
discussed here:
\begin{enumerate}
	\item  Pion HBT does not yield reliable constraints on 
	kinetic (thermal) freeze-out.\footnote{It may 
	be that a treatment of collective flow influence on 
	interferometry using correct statistical distributions can yield a 
	consistent picture for pions and protons and can serve to obtain 
	freeze-out constraints -- these will likely be considerably 
	different from the ones presently used.}
	It may actually be that pion interferometry 
	measures a different, later time than 
	proton correlations and momentum spectra. This late ``wave function 
	freeze-out'' may be related e.g. to very soft or mean field
	interactions of pions which do not alter the momentum distributions 
	as mentioned in \cite{bass}.  
	Momentum spectra, however, favour $T \ge 135 \, \mathrm{MeV}$ 
	and $\left\langle \beta_{T} \right\rangle \le 0.35$.

	\item Freeze-out does not occur with one even approximately 
	universal temperature for different species and also different parts 
	of the source.
\end{enumerate}

%

%
\begin{table}[p]
	\begin{center}
	\caption{Parameters from transverse mass spectra of pions, kaons and 
	protons in central 158$A$GeV Pb+Pb collisions. The slope parameters 
	were taken from \protect\cite{NA44b,na44d,na49stop,na49d}, 
	the average transverse 
	momentum has been calculated assuming an exponential shape of the 
	$m_{T}$ spectra in the fit range used.}
		\begin{tabular}{|cc|c|c|}
			\hline
			 particle & & $T_{\rm eff}$ (MeV) & $\left\langle p_{T} 
			 \right\rangle$ (MeV) \\
			\hline \hline
			$\pi^+$ & NA44 & $156 \pm 6$ & $531$  \\
			\hline
			$\pi^-$ & NA44 & $154 \pm 8$ & $530$  \\
			\hline
			$K^+$ & NA44 & $234 \pm 6$ & $556$  \\
			\hline
			$K^-$ & NA44 & $235 \pm 7$ & $557$  \\
			\hline
			$p$ & NA44 & $289 \pm 7$ & $705$  \\
			\hline
			$\bar{p}$ & NA44 & $278 \pm 9$ & $695$  \\
			\hline
			$p-\bar{p}$ & NA49 & $308 \pm 15$ & $720$  \\
			\hline
			$d$ & NA44 & $376 \pm 17$ & $1115$  \\
			\hline
			$d$ & NA49 & $376 \pm 37$ & $1069$  \\
			\hline
		\end{tabular}
	\label{tbl1}
	\end{center}
\end{table}
%
%
\end{document}